\documentclass[twocolumn]{aastex631}

\usepackage{amsmath}

\usepackage{apjfonts}

\DeclareSymbolFont{cmletters}{OML}{cmm}{m}{it}
\DeclareMathSymbol{v}{\mathalpha}{cmletters}{"76}

\newcommand{\hammer}{H-AMR}

\definecolor{nick}{HTML}{006400}

\shorttitle{BhXRB outbursts}
\shortauthors{Liska et al.}
\graphicspath{{./}{figures/}}

\begin{document}

\title{Magnetic flux plays an important role during a BhXRB outburst in radiative 2T GRMHD simulations}

\correspondingauthor{Matthew Liska}
\email{mliska3@gatech.edu}
\author[0000-0003-4475-9345]{M.T.P. Liska}
\affiliation{Center for Relativistic Astrophysics, Georgia Institute of Technology, Howey Physics Bldg, 837 State St NW, Atlanta, GA 30332, USA}
\affiliation{Institute for Theory and Computation, Harvard University, 60 Garden Street, Cambridge, MA 02138, USA}
\author[0000-0002-5375-8232]{N. Kaaz}
\affiliation{Center for Interdisciplinary Exploration \& Research in Astrophysics (CIERA), Physics \& Astronomy, Northwestern University, Evanston, IL 60202, USA}
\author[0000-0002-2825-3590]{K. Chatterjee}
\affiliation{Black Hole Initiative at Harvard University, 20 Garden Street, Cambridge, MA 02138, USA}
\author[0000-0002-2791-5011]{Razieh Emami}
\affiliation{Institute for Theory and Computation, Harvard University, 60 Garden Street, Cambridge, MA 02138, USA}
\author[0000-0003-1984-189X]{G. Musoke}
\affiliation{Canadian Institute for Theoretical Astrophysics, 60 St George St, Toronto, ON M5S 3H8, Canada}



\begin{abstract}
Black hole (Bh) X-ray binaries cycle through different spectral states of accretion over the course of months to years. Although persistent changes in the Bh mass accretion rate are generally recognized as the most important component of state transitions, it is becoming increasingly evident that magnetic fields play a similarly important role. In this article, we present the first radiative two-temperature (2T) general relativistic magnetohydrodynamics (GRMHD) simulations in which an accretion disk transitions from a quiescent state at an accretion rate of $\dot{M} \sim 10^{-10} \dot{M}_{\rm Edd}$ to a hard-intermediate state at an accretion rate of $\dot{M} \sim 10^{-2} \dot{M}_{\rm Edd}$. This huge parameter space in mass accretion rate is bridged by artificially rescaling the gas density scale of the simulations. We present two jetted BH models with varying degrees of magnetic flux saturation. We demonstrate that in `Standard and Normal Evolution' models, which are unsaturated with magnetic flux, the hot torus collapses into a thin and cold accretion disk when $\dot{M} \gtrsim 5\times 10^{-3} \dot{M}_{\rm Edd}$. On the other hand, in `Magnetically Arrested Disk' models, which are fully saturated with vertical magnetic flux, the plasma remains mostly hot with substructures that condense into cold clumps of gas when $\dot{M} \gtrsim 1 \times 10^{-2} \dot{M}_{\rm Edd}$. This suggests that the spectral signatures observed during state transitions are closely tied to the level of magnetic flux saturation.
\end{abstract}


\section{Introduction}
\label{sec:Intro}
Black hole X-Ray binaries (BhXRBs) are observed in various spectral state of accretion (e.g \citealt{Esin1997, Remillard2006, FenderBelloni2004, Belloni2010, Belloni2016}). Most of the time, BhXRBs are quiescent, with low luminosities and a hard X-Ray spectrum. However, according to our current understanding, thermal-viscous instabilities \citep{LightmanEardley1974, Shakura1976} in the outer accretion disk drive periodic outbursts during which the gas supply falling into the black hole increases by several orders of magnitude (e.g. \citealt{Lasota2001}). When BhXRBs go into outburst, their luminosity increases by several orders of magnitude before and their spectrum changes shape. Similar cycles might exist in AGN (e.g. \citealt{Noda2018}), albeit they occur on much longer timescales and thus cannot be observed. It is also well-known that the (quasi-periodic) variability and jet production are tightly linked to the spectral state of a black hole (e.g. \citealt{IngramMotta2020}). However, due the complexity of the physics involved, such as radiation-matter interaction and thermal decoupling between ions and electrons, our ability to model outbursts is extremely limited. 

Spectral states are classified by two parameters (e.g. \citealt{Kalemci2022}). The first parameter is the luminosity with respect to the Eddington limit ($L_{\rm Edd}=4 \pi GM_{\rm Bh}c / \kappa_{\rm es} \sim 1.3 \times 10^{38} M_{\rm Bh}/M_{\odot}$erg/s, where $\kappa_{\rm es}$ is the opacity due to electron scattering), which represents the point at which radiation pressure overcomes gravity. The second parameter is the hardness of the spectrum, which is typically defined as the luminosity ratio between the hard and soft X-ray bands, and quantifies how much the spectrum deviates from a thermal blackbody. During an outburst, the BhXRB first transitions from the quiescent state to the hard-intermediate state, during which the luminosity rises rapidly and the spectrum remains dominated by hard Comptonized emission. Then, the source transitions to the high-soft state characterised by a soft spectrum and the absence of a jet. After spending (typically) weeks to months in the high-soft state, the source transitions back through several (jetted) intermediate states to the quiescent state. This transition from the high-soft state to the quiescent state occurs at a much lower luminosity than the other way around, for reasons that are not well understood (e.g. \citealt{Begelman2014, Scepi2021}).

Most general relativistic magnetohydrodynamic (GRMHD) simulations to date address accretion in the quiescent state. While BhXRBs indeed spend most of their time in the quiescent state, they accrete most of their gas (and hence grow most rapidly) in the hard-intermediate and high-soft states (e.g. \citealt{Fabian2012}). However, simulating accretion disks in these luminous states is numerically challenging due to the presence of dynamically important radiation fields and thermal decoupling between ions and electrons. Presently, only a handful of GRMHD codes are able to model radiation (e.g. \citealt{Sadowski2013, Mckinney2013, Fragile2014, Ryan2017, White2023}). In addition, since radiative cooling makes such accretion disks thinner, one needs a much higher resolution to resolve them. For example, to resolve on a spherical grid a disk that is two times thinner without static or adaptive mesh refinement requires a factor $32$ more computational time. These factors make such simulations extremely expensive and complex. Due to recent algorithmic and computational advances, radiative GRMHD simulations of accretion disks accreting above a few percent of the Eddington limit (i.e., very thin disks) came within the realm of possibility (e.g. \citealt{Ohsuga2011, Mishra2016, Teixeira2017, Fragile2018, Lancova2019, Mishra2020, Mishra2022, Liska2022, Liska2023}). These recent advances supplement earlier work that attempted to tackle the physics driving accretion in the luminous states using an ad hoc cooling function in place of first-principles radiation (e.g. \citealt{Noble2009, Avara2016, Hogg_Reynolds2017, Hogg_Reynolds2018, Scepi2023, Nemmen2023, Bollimpalli2023}). 

Recently, radiative two-temperature GRMHD simulations of accretion disks accreting at $L \sim 0.35 L_{\rm Edd}$ demonstrated that in systems where no vertical magnetic flux is present, a thin and cold accretion disk forms, possibly explaining the high-soft state \citep{Liska2022}. However, in the presence of dynamically important large scale vertical magnetic flux that saturates the disk, the accretion disk enters the magnetically arrested disk state \citep[e.g. `MAD', ][]{Narayan2003, Tchekhovskoy2011, Mckinney2012}. In MADs magnetic flux impedes smooth inflow of gas, and instead, gas accretion proceeds much more chaotically (e.g. \citealt{Begelman2021}).  Moreover, the typically cold and slender accretion disk undergoes truncation, transforming into a two-phase coronal plasma characterized by the coexistence of cold gas clumps enveloped by a hotter and sparser gas environment within a (truncation) radius of approximately $r \lesssim 20 r_g$ \citep{Liska2022}. These magnetically truncated disks provide a promising model for the luminous-hard state. First, the decoupling of the plasma into a two-phase medium might explain why hard X-Ray emission dominates over the thermal blackbody emission from the disk. Second, the presence of cold plasma down to the innermost stable circular orbit (ISCO) provides an interesting explanation for the observed relativistic broadened iron-reflection lines in the luminous-hard state (e.g. \citealt{Reis2010}), which before this work was thought to only feature hot gas unable to produce such lines. In between the zero-magnetic flux and MAD regime, where vertical magnetic flux is present but does not saturate the disk, \citet{Lancova2019} demonstrated that a hot plasma with both inflowing and outflowing components sandwiches a cold and thin accretion disk. Such puffy disk models can potentially describe BhXRBs in some intermediate spectral state (e.g. between the luminous-hard and high-soft states), which launch relativistic jets but show no clear evidence of significant disk truncation (e.g. \citealt{Kara2019}).

However, none of the previously discussed work addresses at which accretion rate a hot quiescent-state torus transitions to a thin (truncated) hard-intermediate state disk and what role magnetic fields play in that process. In this article we present the first radiative two-temperature general relativistic magnetohydrodynamics (GRMHD) simulations spanning 8 orders of magnitude in mass accretion rate. These simulations demonstrate a transition from a hot torus in the quiescent state to either a magnetically truncated (e.g. \citealt{Liska2022}) or puffy accretion disk (e.g. \citealt{Lancova2019}) in the (hard-) intermediate state depending on the amount of magnetic flux saturation. In Section \ref{sec:Physical_Setup} we describe our radiative two-temperature GRMHD code and numerical setup, before presenting our results in Section \ref{sec:Results} and concluding in Section \ref{sec:discussion}.

\section{Numerical Setup}
\label{sec:Numerical_Setup}
To model the rise from quiescence to the hard-intermediate state we use the GPU-accelerated GRMHD code \hammer{} \citep{Liska2018A, Liska2020}. \hammer{} evolves the radiative two-temperature GRMHD equations (e.g. \citealt{Sadowski2013, Sadowski2017}) on a spherical grid in Kerr-Schild coordinates. Similar to \citealt{Liska2022}, we model the radiation field as a second fluid using the M1 approximation and, in addition, also evolve the photon number density to get a better estimate for the radiation temperature \citep{Sadowski2015}. Radiative processes such as Brehmstrahlung, Synchrotron, bound-free, iron line emission, and scattering (including Comptonization) are included assuming a $M_{\rm Bh}=10 M_{\odot}$ black hole surrounded by plasma with solar abundances ($X=0.70$, $Y=0.28$, $Z=0.02$). The associated energy-averaged grey opacities are provided in \citet{Mckinney2017} (equations C16, D7 and E1). 

At each timestep, the dissipation rate is calculated by subtracting the internal energy provided by the entropy equation from the internal energy provided by the energy equation (e.g. \citealt{Ressler2015}). Subsequently, the total energy dissipation is divided as a source term between the electron and ions based on a reconnection heating model \citep{Rowan2017}. This deposits a fraction $\delta_e \lesssim 0.5$ of the dissipation into the electrons, which varies between $\delta_e \sim 0.2$ in less magnetized regions to $\delta_e \sim 0.5$ in highly magnetized regions. Coulomb collisions \citep{Stepney1983} are taken into account through an implicit source term  (e.g. \citealt{Sadowski2017}). To avoid the jet funnel becoming devoid of gas and to keep the GRMHD scheme stable, we floor the density in the drift frame of the jet \citep{Ressler2015} such that the ratio of the density and magnetic pressure $\frac{p_{b}}{\rho c} \lesssim 12.5$. 

We use a spherical grid in the Kerr-Schild foliation with coordinates $x^1=\log(r)$, $x^2=\theta$, and $x^3=\varphi$ with a resolution of $N_r \times N_{\theta} \times N_{\varphi}=420\times192\times192$ for our SANE model and $N_r \times N_{\theta} \times N_{\varphi}=560\times192\times192$ for our MAD model. This adequately resolves the fastest growing MRI-wavelength by $\gtrsim 16$ cells in all 3 dimensions. We place the outer boundary at $R_{\rm out}=10^3 r_g$ for our SANE model, and at $R_{\rm out}=10^4 r_g$ for our MAD model. We also maintain at least 5 cells within the event horizon such that the inner boundary is causally disconnected from the rest of the computational domain. We speed up the simulations approximately $3$-fold by introducing $4$ levels of local adaptive timestepping \citep{Liska2020}. To prevent cell squeezing around the polar axis from slowing down our simulations (e.g. \citealt{Courant1928}) we use $4$ levels of static mesh derefinement \citep{Liska2018A, Liska2020} to reduce the $\varphi-$resolution to $N_{\varphi}=[96,48,24,12]$ within $\theta \lesssim [30^{\circ}, 15^{\circ}, 7.5^{\circ}, 3.75^{\circ}]$ from each pole. This maintains a cell aspect ratio of roughly $|\Delta r|:|\Delta \theta|:|\Delta \varphi| \sim 1:1:2$ throughout the grid, which is sufficient to capture the 3-dimensional nature of the turbulence. 

\begin{figure}
    \centering
    \includegraphics[clip,trim=0.0cm 0.0cm 0.0cm 0.0cm,width=\columnwidth]{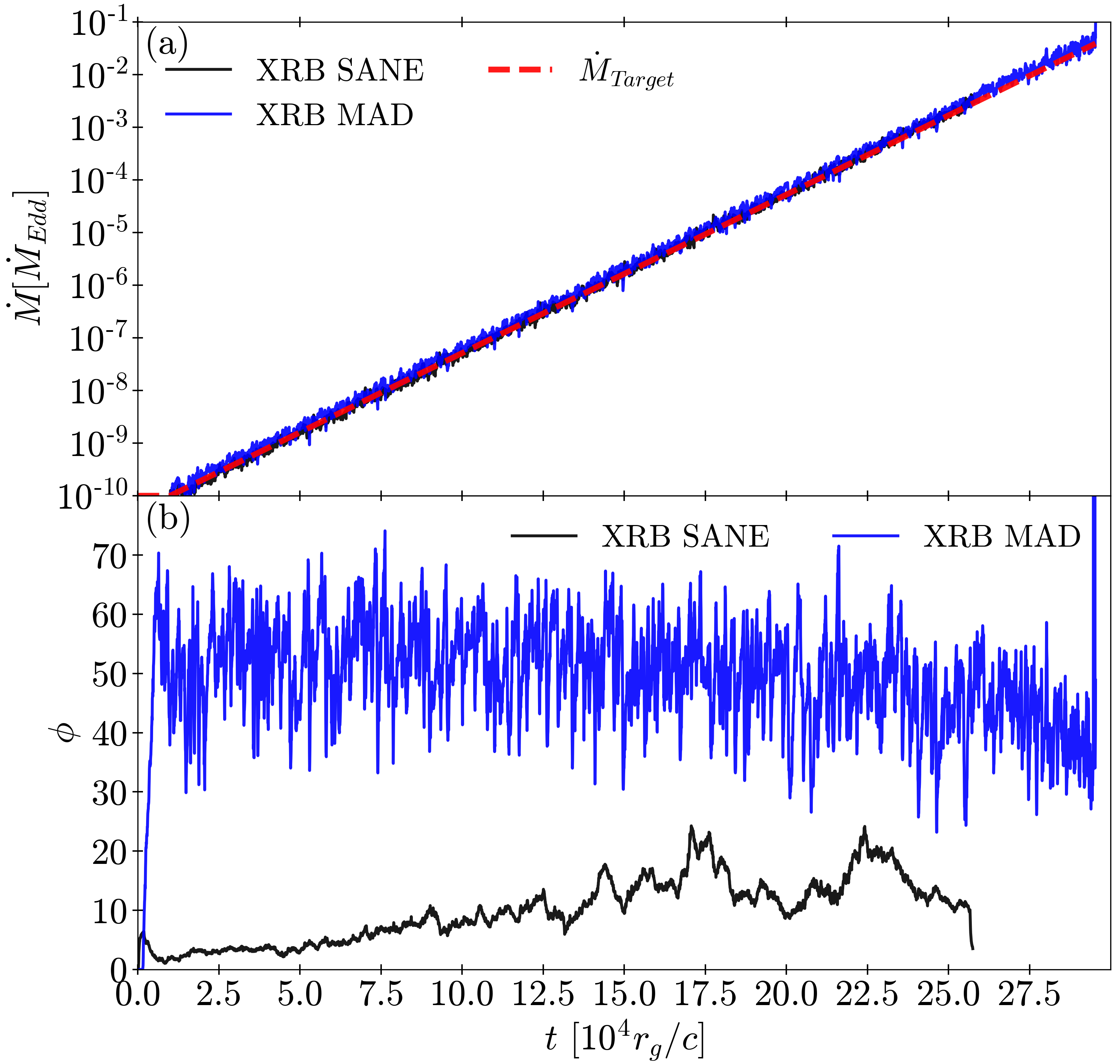}
    \vspace{-10pt}
    \caption{\textbf{Panel A:} The event horizon mass accretion rate $\dot{M}$ closely follows the target mass accretion rate $\dot{M}_{\rm target}$ (red) for both the SANE (black) and MAD (blue) models. \textbf{Panel B: } The normalized magnetic flux $\phi$ maintains saturation value in the MAD model and stays a factor $\gtrsim 2.0$ below saturation in the SANE model.}
    \label{fig:timeplot}
\end{figure}

\begin{figure*}
    \centering
    \includegraphics[width=1.0\linewidth,trim=0mm 0mm 0mm 0,clip]{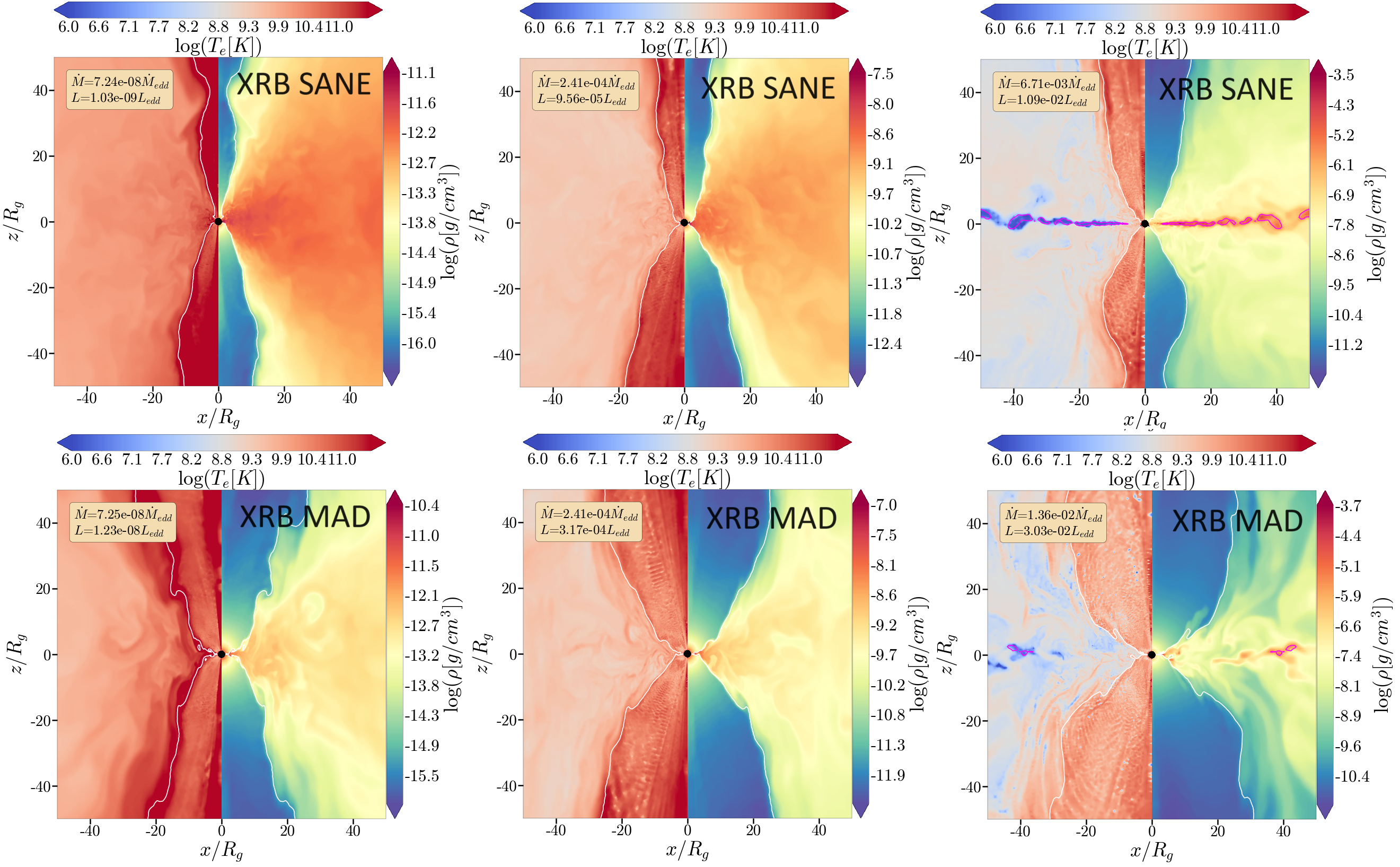}
    \caption{The SANE (upper panels) and MAD (lower panels) models at 3 different accretion rates. The left hemisphere illustrates the electron temperature ($T_{e}$) while the right hemisphere illustrates the density ($\rho$). The disk-jet boundary ($b^2/(\rho c^2)=1$) is demarcated by white line and the last scattering surface ($\tau_{es}=1$) is demarcated by a magenta line. The inset in the left hemisphere gives the mass accretion rate and luminosity in Eddington units. See our \href{https://www.youtube.com/playlist?list=PLDO1oeU33Gwm1Thyw0iHC14BbvBWaG5cE}{Youtube playlist} and the supplementary materials for an animation of the density, electron temperature, and ion temperature for both models. At very low accretion rates (left panels), the electron temperature is determined by the heating rate and adiabatic evolution of the electrons. At intermediate accretion rates (middle panels), the electrons cool efficiently but there is no noticeable change in the disk structure. At accretion rates of $\dot{M} \gtrsim 10^{-2} \dot{M}_{\rm Edd}$ the torus collapses into a thin accretion disk (SANE) sandwiched by a hot plasma or forms a magnetically truncated accretion disk (MAD).}
    \label{fig:contourplot}
\end{figure*}

\section{Physical Setup}
\label{sec:Physical_Setup}
To understand the effects of magnetic flux saturation on the transition from the quiescent to the hard-intermediate state, we include two models in the SANE \citep[`Standard and Normal Evolution', ][]{Narayan1994} and MAD \citep[`Magnetically Arrested Disk', ][]{Narayan2003} regimes. We assume a rapidly spinning black hole with spin parameter $a=0.9375$. Our SANE model (XRB SANE) features a standard Fishbone \& Moncrief torus \citep{Fishbone1976} with inner radius $r_{\rm in}=6\,r_g$, radius of maximum pressure at $r_{\rm max}=12\,r_g$, and outer radius $r_{\rm out} \sim 50 r_g$. Our MAD model (XRB MAD), on the other hand, features a much larger torus with $r_{\rm in}=20\,r_g$ and $r_{\rm max}=41 r_g$ whose outer edge lies at $r_{\rm out} \sim 800\,r_g$. These torii are pretty standard choices in the GRMHD community (e.g. \citealt{Porth2019, Chatterjee2023}). We thread the SANE model with magnetic vector potential,
\begin{equation}
A_{\Phi} \propto \rho-0.2, 
\end{equation}
and the MAD model with magnetic vector potential, 
\begin{equation}
A_{\Phi} \propto \rho r^3 \sin^3 \left(\theta \right) \exp \left(-\frac{r}{400\,r_g}\right) - 0.2, 
\end{equation}
where $\rho$ is the gas density. In both cases this produces a poloidal magnetic flux loop that is approximately the size of the torus. Since the MAD torus is much larger than the SANE torus, only the MAD torus contains sufficient magnetic flux to enter the MAD state. There is no toroidal magnetic flux present in the initial conditions. In both cases, we normalize the resulting magnetic field such that $\beta^{\rm max}=p^{\rm max}_{\rm gas}/p^{\rm max}_{b}=100$ where $p^{\rm max}_{\rm gas}$ and $p^{\rm max}_{b}$ are the maximum gas and magnetic pressure in the torus. For the purpose of calculating the initial torus solution we set the adiabatic index $\gamma=5/3$ for our SANE model and $\gamma=13/9$ for our MAD model. We subsequently distribute, according to our heating prescription involving a magnetic reconnection model \citep{Rowan2017}, the total pressure between the ions and electrons before we self-consistently evolve their entropy and adiabatic indices (e.g. \citealt{Sadowski2017}). 

To make GRMHD simulations of BhXRB outbursts feasible, we artificially shorten the relevant timescales by introducing a rescaling method that, as a function of time, sets the accretion rate to a predetermined value. However, before we apply this method, we first run the simulation for $t=10^4 r_g/c$ in two-temperature non-radiative GRMHD to get the accretion disk into a quasi-steady state. We subsequently restart the simulation in radiative two-temperature GRMHD and re-normalize the density ($\rho$) every timestep ($\Delta t \sim 10^{-2} r_g/c$) with a factor $\zeta$ such that the running average of the mass accretion rate over $\Delta t =10^3 r_g/c$ at $r=5r_g$,
\begin{equation}
\langle \dot{M}_{5 r_g} \rangle= \langle [\int -\sqrt{-g} \rho u^r d\theta d\varphi]_{r=5 r_g} \rangle,
\end{equation}
 is scaled to a time-dependent `target' mass accretion rate,
 \begin{equation}
\dot{M}_{\rm Target}=10^{-10}\times2^{\frac{t-10^4}{10^4}} \dot{M}_{\rm Edd},
 \end{equation}
 via the rescaling factor,
 \begin{equation}
     \zeta = \frac{\dot{M}_{\rm Target}}{\langle \dot{M}_{5r_g} \rangle},
 \end{equation}
where $\dot M_{\rm Edd}=L_{\rm Edd} / (\eta_{\rm NT}c^2)$ is the Eddington accretion rate and $\eta_{\rm NT}=0.178$ is the analytically derived radiative efficiency for a radiatively efficient thin accretion disk \citep{Novikov1973}. In addition to the density ($\rho$), we also rescale the internal energy density ($u_g$), radiation energy density ($E_{\rm rad}$) and magnetic energy density ($b^2$) with the same prefactor $\zeta$. This leads to a doubling of the black hole mass accretion rate every $t=10^4 r_g/c$. Note that this approach automatically increases the total amount magnetic flux at the event horizon $r_{\rm Bh}$, 
\begin{equation}
\Phi_{r_{\rm Bh}}=\frac{\sqrt{4\pi}}{2} [\int \sqrt{-g} |B^r| d\theta d\varphi]_{r=r_{\rm Bh}}, 
\end{equation}
by a factor $\sqrt{\zeta}$, such that the normalized magnetic flux,
\begin{equation}
\phi_{r_{\rm Bh}} =\frac{\Phi_{r_{\rm Bh}}}{\sqrt{\langle \dot{M}_{5r_g} \rangle}}, 
\end{equation}
remains constant. When $\phi \sim 40-50$ we expect that the disk turns MAD and flux bundles get ejected from the black hole (e.g. \citealt{Tchekhovskoy2011, Mckinney2012}). We achieve inflow-outflow equilibrium over the mass accretion rate doubling time ($\Delta t = 10^{4} r_g/c$) up to approximately $r \sim 15 r_g$ in our SANE model and $r \sim 30 r_g$ in our MAD model. This is the radius within which our models are expected to converge to their steady accretion rate analogues with a constant $\zeta$.

Note that constant-$\zeta$ GRMHD simulations typically exhibit a factor $\sim 5$ variability in the dimensionless accretion rate during the first $t \sim 10,000 r_g/c$ when seeded from a torus that has not yet fully developed MRI turbulence (e.g. \citealt{Porth2019}). Thus, to span the given range in $\dot{M}$, we argue that rescaling $\dot{M}$ in a single ultra-long duration GRMHD model is superior to running a suite of $\sim 25-30$ GRMHD models with a constant $\zeta$ that each start from an equilibrium torus and only last for a duration of $\Delta t=10,000 r_g/c$. Namely, the variability of $\dot{M}$ (Fig.~\ref{fig:timeplot}) during each $\Delta t = 10,000 r_g/c$ time interval (after $t=10,000 r_g/c$) is at most a factor $\lesssim 3$ which is less than the variability during the first $\Delta t=10,000 r_g/c$. In other words, we argue that using a well converged turbulent accretion state at, for example, an accretion rate $\dot{M}=1-2 \times 10^{-5} \dot{M}_{\rm Edd}$ as the initial conditions for a GRMHD model with an accretion rate $\dot{M}=2-4 \times 10^{-5} \dot{M}_{\rm Edd}$ is preferred over restarting the entire simulation from scratch.

\section{Results}
\label{sec:Results}
We evolve both models for $t \sim 260,000 - 280,000 \,r_g/c$, during which the targeted mass accretion rate increases by $8$ orders of magnitude. As illustrated in Figure \ref{fig:timeplot}a, the black hole mass accretion rate closely follows the targeted mass accretion rate for both models. However, as illustrated in Figure \ref{fig:timeplot}b, while the normalized magnetic flux threading the Bh event horizon stays constant in our MAD model, it increases by a factor of $\sim 3$ in our SANE model. This is because a significant fraction of the initial gas reservoir accretes or gets ejected in the form of winds, leading to a relative larger increase in the dimensionless magnetic flux compared to the mass accretion rate. The rapid variability of the magnetic flux observed in our MAD model is a well-known characteristic of MADs (e.g. \citealt{Tchekhovskoy2011, Mckinney2012}) caused by flux bundles being ejected from the black hole event horizon through magnetic reconnection (e.g. \citealt{RipperdaLiskaEtAl2021}). 

The contour plots of density and electron temperature in Figure \ref{fig:contourplot} illustrate 3 different stages of the artificially-induced state transition. An accompanying animation also illustrating the ion temperature is included in the supplementary materials and on our \href{https://www.youtube.com/playlist?list=PLDO1oeU33Gwm1Thyw0iHC14BbvBWaG5cE}{Youtube playlist}. In the first stage ($\dot{M} \lesssim 10^{-6}\dot{M}_{\rm Edd}$), the radiative efficiency is low and radiative cooling plays a negligible role. The ions in the plasma are significantly hotter than the electrons because our heating prescription typically injects only a fraction $\delta_e \sim 0.2-0.4$ of the dissipative heating into the electrons (and a fraction $\delta_i \sim 0.6-0.8$ into the ions). In the second stage ($\dot{M} \gtrsim 10^{-6}\dot{M}_{\rm Edd}$), radiative cooling of the electrons becomes efficient leading to a drop in electron temperature but no other structural change (see also \citealt{Chatterjee2023}). In the third stage ($\dot{M} \gtrsim 10^{-2}\dot{M}_{\rm Edd}$), Coulomb collisions become efficient. This allows the ions to cool by transferring their energy to the radiation-emitting electrons and, eventually, leads to a rapid collapse of the hot torus. 

\begin{figure}
    \centering
    \includegraphics[clip,trim=0.0cm 0.0cm 0.0cm 0.0cm,width=\columnwidth]{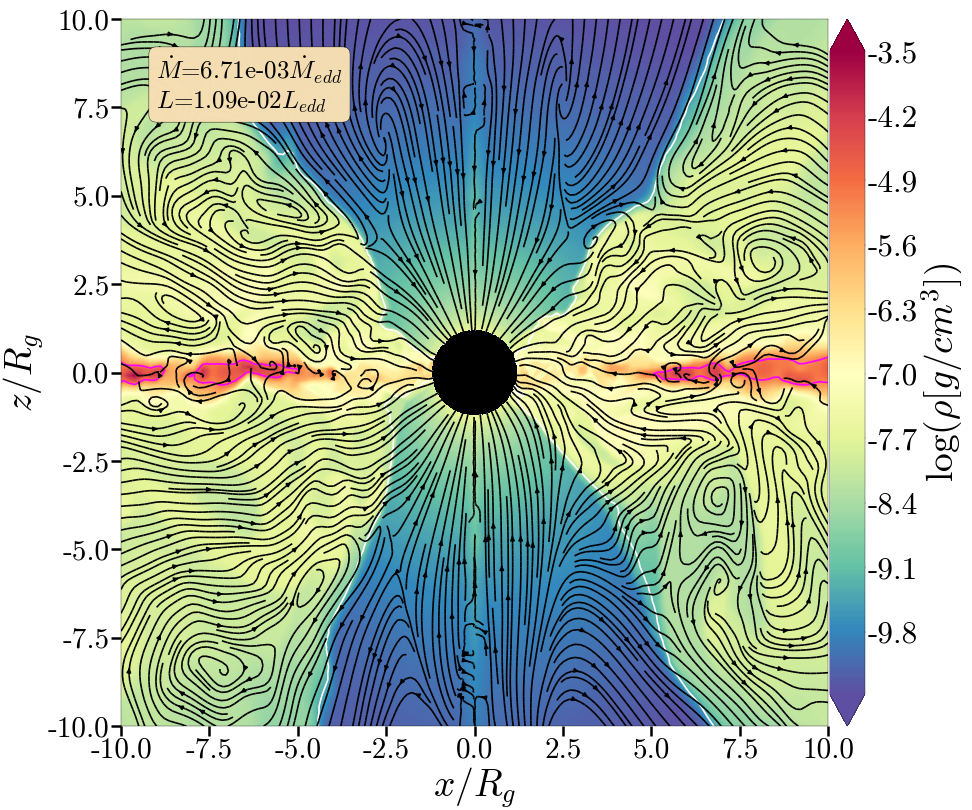}
    \vspace{-10pt}
    \caption{A contourplot of density with velocity streamlines in black and the last scattering surface in magenta for model XRB SANE. Similar to the puffy accretion disk models presented in \citet{Lancova2019} the majority of gas accretion seems to be driven by inflows outside of the disk's midplane.}
    \label{fig:streamplot_XRB}
\end{figure}

In our SANE model this collapse results in a geometrically thin accretion disk surrounded by hot magnetic pressure supported gas outside of $r \gtrsim 3\,r_g$. Thus, the disk is only truncated very close to the black hole. The production of hot non-thermal electrons within the ISCO was predicted by \citet{Hankla2022}. Interestingly, this hot coronal gas rather than the thin accretion disk seems to be responsible for the majority of $\dot{M}$ (Fig.~\ref{fig:streamplot_XRB}). This is similar to the puffy accretion disks first presented in other radiative GRMHD simulations \citep{Lancova2019} and in pure MHD simulations of weakly to moderately magnetized disks \citep{jacquemin_2021}. On the other hand, in our MAD model the hot torus transitions into a two-phase medium with cold optically thick patches of gas surrounded by hot, optically thin, plasma. These cold patches of gas are visible for $20\,r_g \lesssim r \lesssim 100\,r_g$  and do not reach the event horizon. Since this work was performed at a rather low resolution, we were forced to terminate both our SANE and MAD models around $\dot{M} \sim 10^{-2} \dot{M}_{\rm Edd}$ because the scaleheight of the disk dropped below $h/r \lesssim 0.1$ and the cold plasma became under-resolved. Nevertheless this work addresses at which accretion rate the hot torus transitions into a (truncated) cold accretion disk.  Future work featuring at least an order of magnitude higher effective resolution will be necessary to understand how the disk further evolves as we keep increasing the mass accretion rate.

These findings diverge from the magnetically truncated accretion disk models detailed in \citet{Liska2022}, where a slender disk was truncated at $r \sim 20 r_g$ and cold patches of gas reached the event horizon. The absence of this thin disk in our simulations may be attributed to the considerably higher saturation of magnetic flux within our torus, distinguishing it from the disk in \citealt{Liska2022}. This discrepancy could feasibly result in a significantly larger truncation radius. Consequently, if the truncation radius in our MAD model lies much further out, it is plausible that our simulation's duration is insufficient to capture the formation of a thin accretion disk. A larger truncation radius (assuming the cold clumps of gas were sufficiently resolved, which might not be true) might consequently also explain why no cold plasma reaches the event horizon. Namely, as proposed in \citet{Liska2022}, magnetic reconnection can potentially evaporate the cold clumps of gas before they reach the event horizon. This is less likely to happen if the magnetic truncation radius moves further in and, hence, the cold clumps have less time to evaporate. 

\begin{figure*}
    \centering
    \includegraphics[clip,trim=0.0cm 0.0cm 0.0cm 0.0cm,width=0.605\linewidth]{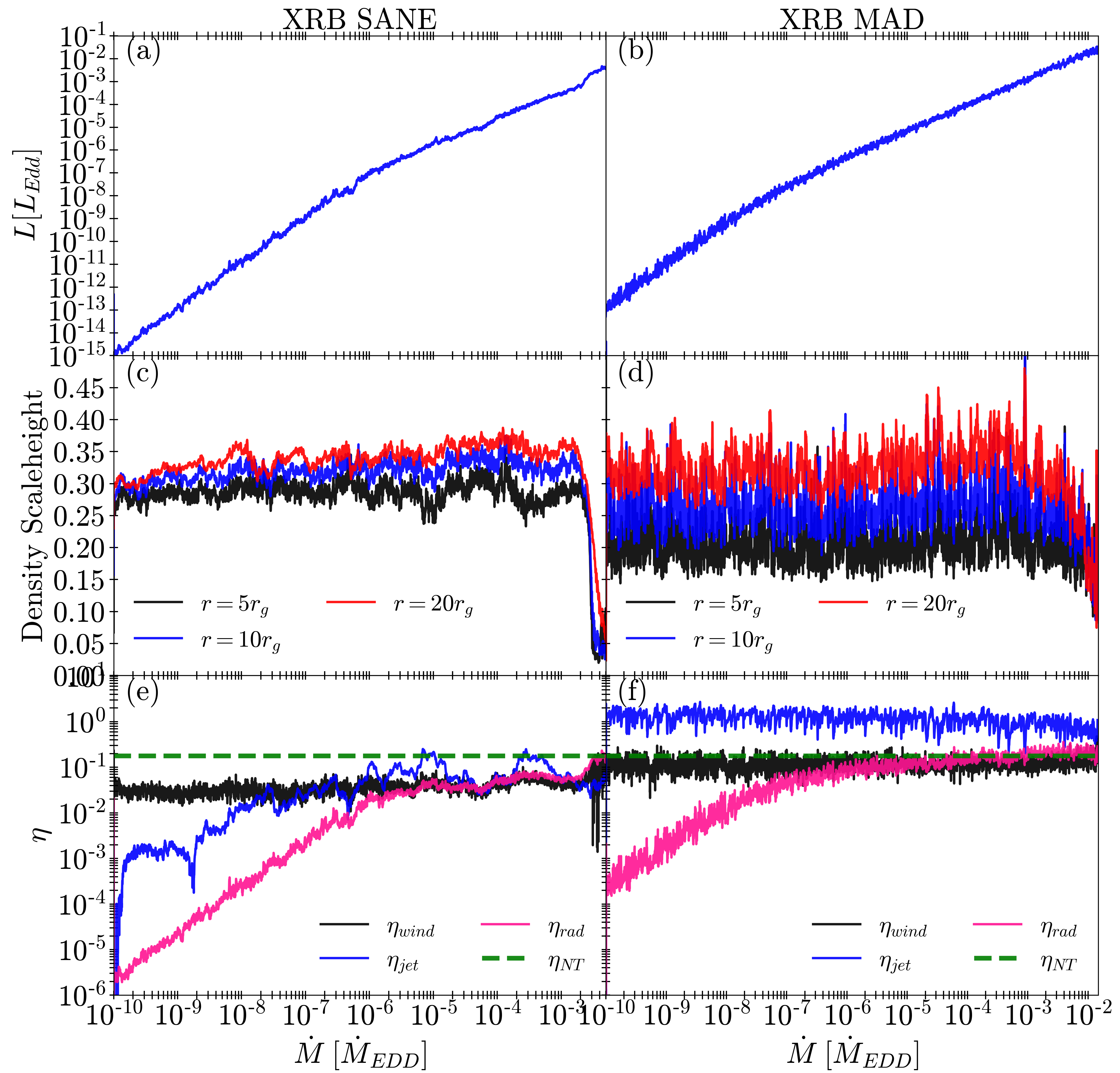}
    \vspace{-9pt}
    \caption{\textbf{Panels (a,b):} As the mass accretion rate rises the luminosity increases from $L \sim 10^{-15}-10^{-13} L_{EDD}$ to $L \sim 10^{-2} L_{EDD}$. This is driven by both an increase in the mass accretion rate and radiative efficiency. \textbf{Panels (c,d):} The density scaleheight of the disk stays relatively constant until the accretion rate exceeds $\dot{M} \sim 10^{-3} \dot{M}_{EDD}$ at which point the disk collapses. At this point Coulomb collisions are efficient and both the ions can cool through the radiation emitting electrons. \textbf{Panels (e,f):} The jet (blue), wind (black), radiative (purple) and NT73 (dashed green) efficiency as function of $\dot{M}$. Interestingly, the MAD model maintains a significantly higher radiative efficiency, presumably due to more efficient Synchrotron emission.}
    \label{fig:mdotplot}
\end{figure*}

In Figure \ref{fig:mdotplot} we plot the time evolution of the bolometric luminosity (panels a and b), density scale height (panels c and d), and outflow efficiencies (panel c and d) as function of the mass accretion rate. While the luminosity increases from $L=10^{-15} L_{\rm Edd}$ to $L=10^{-2} L_{\rm Edd}$, the radiative efficiency increases by $3-5$ orders of magnitude. Similar to results presented in the radiative GRMHD simulations of \citet{Ryan2017} and \citet{Dexter2021}, the MAD model is significantly more radiatively efficient, especially at low accretion rates. This is caused by more efficient Synchrotron cooling in the highly magnetized gas of a MAD. Around $\dot{M}=5 \times 10^{-3} \dot{M}_{\rm Edd}$ the SANE model collapses into a thin accretion disk, and we observe a rapid order-of-magnitude rise in the radiative efficiency to the NT73 \citep{Novikov1973} limit of $\eta_{rad} \sim \eta_{NT} \sim 0.18$. Here,
\begin{equation}
\eta_{rad}=\frac{[\int \sqrt{-g} R^{r}_{t} d\theta d\varphi]_{r=200 r_g}}{\langle \dot{M}_{5 r_g}\rangle},
\end{equation}
with $R^{\mu}_{\nu}$ being the radiation stress-energy tensor. This collapse manifests itself as a rapid decrease in the density scale height of the disk,
\begin{equation}
\frac{h}{r}=\langle\langle |\theta - \pi/2 |\rangle\rangle_{\rho}, 
\end{equation}
with $\langle\langle x \rangle \rangle_{\rho}$ denoting a density-weighted average. On the other hand, in our MAD model the radiative efficiency asymptotes to $\eta_{rad} \sim 1.2 \eta_{NT}$. This has been observed in other radiative MADs (e.g. \citealt{Curd2023}) and could, pending future analysis, potentially be explained by the presence of a dynamically important magnetic field that injects energy into the accreting gas, which is not accounted for in \citet{Novikov1973}. In addition, there is only a marginal factor $\sim 2$ decrease in the disk scale height after the formation of cold plasma, because magnetic pressure is able to stabilize the accretion disk against runaway thermal collapse  (e.g \citealt{Sadowski2016_mag, Jiang2019}). To better understand the outflows we subdivide the plasma into jet and wind components at the $b^2/(\rho c^2) = 1$ boundary. Using this definition we can define the wind and jet driven outflow efficiencies in terms of total energy accretion rate $\dot{E}$ and energy accretion rate in the jet $\dot{E}^{jet}$:
\begin{equation}
\eta_{wind} = \frac{ \dot{M}_{5 r_g} -\dot{E}_{5 r_g} + \dot{E}^{jet}_{5r_g}}{\langle \dot{M}_{5 r_g} \rangle}
\end{equation}
\begin{equation}
\eta_{jet}= \frac{- \dot{E}^{jet}_{100r_g} }{\langle \dot{M}_{5r_g} \rangle},
\end{equation}
While both the wind and jet efficiency remain constant within a factor unity in our MAD model, the increase of the normalized magnetic flux in our SANE model causes the jet to become significantly more efficient over time (e.g. $\eta_{jet} \propto \phi^2$).

\begin{figure*}
    \centering
    \includegraphics[clip,trim=0.0cm 0.0cm 0.0cm 0.0cm,width=0.605\linewidth]{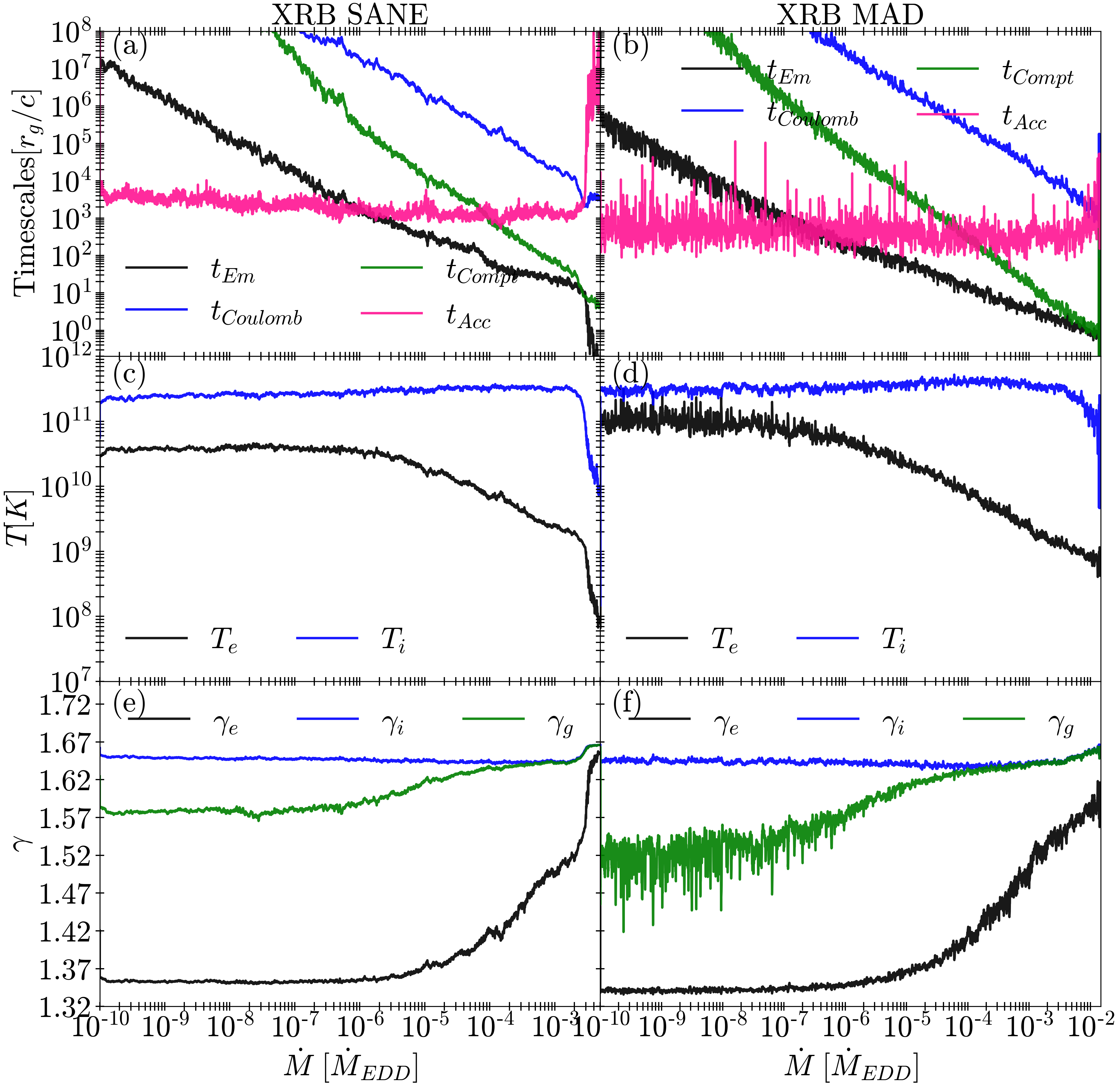}
    \vspace{-9pt}
    \caption{\textbf{Panels (a,b):} In both models \textbf{at $r=10 r_g$} the electron temperature $T_e$ drops for $\dot{M} \gtrsim 10^{-6} \dot{M}_{EDD}$ due to more efficient radiative cooling. The ion temperature $T_i$ is governed by the ability of the ions to transfer their energy to the radiation-emitting electrons and doesn't start cooling until $\dot{M} \sim 10^{-3} \dot{M}_{EDD}$. \textbf{Panels (c,d)}: We compare at $r=10r_g$ the radiative emission ($t_{Em}$), Comptonization ($t_{Compt}$), and Coulomb collision ($t_{Coulomb}$) timescale against the accretion ($t_{acc}$) timescale to illustrate the importance of these processes at different accretion rates. \textbf{Panels (e,f)}: The adiabatic index at $r=10 r_g$ for the electrons $\gamma_e$ is relativistic while the adiabatic index for the ions $\gamma_{i}$ is non-relativistic. This leads to a semi-relativistic gas with adiabatic index $\gamma_{g} \sim 1.55$ at $\dot{M} \lesssim 10^{-6} \dot{M}_{\rm Edd}$.}
    \label{fig:coolplot}
\end{figure*}

To better understand when, during an outburst, certain physical processes become important, we plot in Figure \ref{fig:coolplot}(a,b) the radiative cooling timescale,
\begin{equation}
t_{rad}=\frac{[\int \sqrt{-g} u_e d\theta d\varphi]_{r= 10r_g}}{[\Lambda_{Em}]_{r= 10r_g}},
\end{equation}
with $\Lambda_{Em}$ the radiative emission rate and $u_{i,e}$ the ion/electron internal energy, the Comptonization timescale, 
\begin{equation}
t_{Compt}=\frac{[\int \sqrt{-g} u_e d\theta d\varphi]_{r= 10r_g}}{[\Lambda_{Compt}]_{r= 10r_g}},
\end{equation}
with $\Lambda_{Compt}$ the Compton scattering emission rate, Coulomb coupling timescale, 
\begin{equation}
t_{Coulomb}=\frac{[\int \sqrt{g} \left(u_e + u_i\right) d\theta d\varphi]_{r= 10r_g}}{[\Lambda_{Coulomb}]_{r= 10r_g}},
\end{equation}
with $\Lambda_{Coulomb}$ the Coulomb coupling rate, and the accretion timescale, 
\begin{equation}
t_{Acc}=\frac{[\int \sqrt{-g} \rho r d\theta d\varphi]_{r= 10r_g}}{[\int \sqrt{-g} \rho u^r d\theta d\varphi]_{r= 10r_g}}.
\end{equation}
$\Lambda_{\rm Em}$, $\Lambda_{\rm Compt}$, and $\Lambda_{\rm Coul}$ are derived from the opacities given in \citet{Mckinney2017} and the Coulomb coupling rate given in \citet{Sadowski2017}. Evidently, the radiative and Compton cooling timescale become similar to the accretion timescale around $\dot{M} \gtrsim 10^{-6.5} \dot{M}_{\rm Edd}$ and $\dot{M} \gtrsim 10^{-4} \dot{M}_{\rm Edd}$ respectively. This manifests itself in figure \ref{fig:coolplot}(c,d) as a decrease in the electron temperature,
\begin{equation}
T_e=\frac{[\int \sqrt{-g}  p_e  d\theta d\varphi]_{r= 10r_g}}{[\int \sqrt{-g} \rho d\theta d\varphi]_{r= 10r_g}}.
\end{equation}
On the other hand, the ion temperature,
\begin{equation}
T_i=\frac{[\int \sqrt{-g}  p_i  d\theta d\varphi]_{r= 10r_g}}{[\int \sqrt{-g} \rho d\theta d\varphi]_{r= 10r_g}},
\end{equation}
only drops when the Coulomb coupling timescale becomes comparable to the accretion timescale around $\dot{M} \sim 5 \times 10^{-3} \dot{M}_{\rm Edd}$. Meanwhile the plasma transitions in figure \ref{fig:coolplot}(e,f) from a quasi-relativistic adiabatic index $\gamma \sim 1.5$ to a non-relativistic $\gamma\sim 5/3$. Even at accretion rates that are typically associated with radiatively inefficient accretion ($\dot{M} \lesssim 10^{-7} \dot{M}_{\rm Edd}$), future work will need to test if electron cooling (see also \citealt{Dibi2012, Yoon2020}) and/or a self-consistent adiabatic index can change the spectral signatures compared to equivalent non-radiative single-temperature GRMHD models (e.g. \citealt{Moscibrodzka2014}).

\section{Discussion and Conclusion}
\label{sec:discussion}
In this article we addressed for the first time the transition from the quiescent to the hard intermediate state using radiative two-temperature GRMHD simulations. By rescaling the black hole mass accretion rate across 8 orders of magnitude, these simulations demonstrated that radiative cooling and Coulomb coupling become increasingly important and eventually lead to a transition to a two-phase medium. While the hot torus in SANE models transitions to a thin accretion disk surrounded by a sandwich-like corona above $\dot{M} \gtrsim 5 \times 10^{-3} \dot{M}_{\rm Edd}$, reminiscent of a puffy \citep{Lancova2019} or magnetically elevated \citep{Begelman2017} disk, the MAD torus transitions to a two-phase medium with cold clumps of gas embedded in a hot corona (see also \citealt{Bambic2023}) above $\dot{M} \gtrsim 1 \times 10^{-2} \dot{M}_{\rm Edd}$. The formation of a two-phase plasma consisting out of cold clumps surrounded by hot gas was already demonstrated in our previous work \citep{Liska2018C, Liska2022}. However, while in \citet{Liska2022} a cold accretion disk was present beyond $r \gtrsim 20 r_g$, no cold accretion disk forms in our MAD models. The absence of a truncated disk could potentially be explained by the amount and location of the excess magnetic flux that does not thread the event horizon. Namely, while in \citet{Liska2022} the magnetic flux peaked around $r_t \sim 20-25 r_g$, which is coincident with the magnetic truncation radius, the magnetic flux in this work does not peak until $r_t \sim 40 r_g$. This suggests that the torus is MAD much further out, and thus the magnetic truncation radius is larger than in \citet{Liska2022}. Since the timescale for the disks to come into inflow equilibrium at $r_t \sim 40 r_g$ exceeds $t \gtrsim 20,000 r_g/c$ it is conceivable that a truncated disk did not have sufficient time to form in our MAD models. Future work will need to address how the magnetic truncation radius evolves as the disk gets more saturated with vertical magnetic flux. 

We expect that the MAD and SANE models will exhibit markedly distinct spectral and temporal variability characteristics. For instance, MAD models are projected to give rise to truncated accretion disks, which may correspond to high-luminosity spectral states displaying evidence of disk truncation. Even though cold clumps of gas can break of the inner disk and reach the event horizon as demonstrated in \citet{Liska2022}, the thermal emission will be suppressed due to the much smaller coverage fraction in magnetically truncated disks. Consequently, we anticipate the emission to be predominantly governed by the hot coronal plasma. Conversely, in SANE models, we anticipate a considerably more pronounced thermal spectrum emanating from the accretion disk as it extends down to the innermost stable circular orbit (ISCO). The magnitude of hard X-ray emission will hinge upon optical depth of the coronal plasma. As suggested by \citet{Liska2022}, accretion disks devoid of vertical magnetic flux are improbable to generate adequate hot plasma for producing hard X-ray emission. Such models with a net zero magnetic flux are likely pertinent to the high-soft state. Nevertheless, in SANE models featuring some vertical magnetic flux, puffy accretion disks can form \citep{Lancova2019}, where hot plasma sandwiches the accretion disk and (potentially) hardens the (otherwise) soft disk spectrum.

The goal of this article was to study the transition from the quiescent state to the hard intermediate state, both of which feature radio jets. Thus, we have not considered models that do not produce any jets such as models with a purely toroidal field (e.g. \citealt{Liska2022}) and instead only considered a jetted SANE model with $\phi \sim 5-15$ and a jetted MAD model with $\phi \sim 40-55$. By rescaling both the gas density and magnetic energy density proportionally to the target mass accretion rate, this work implicitly assumes that the accretion-rate normalized magnetic flux ($\phi$) remains constant within a factor $\sim 2$. Conventional thinking might suggest that since the hard-intermediate state is associated with the most powerful jets, and recent polarimetric Event Horizon Telescope observations  \citep{EHT_POLARIZATION} strongly imply that AGN accretion disks in the quiescent state are MAD, the hard-intermediate state would be MAD as well. However, the jet power is set by the total amount of magnetic flux threading the black hole ($P_{\rm jet} \propto \Phi^{2}$), and thus a SANE jet at a much higher accretion rate can easily outperform a MAD jet at a lower accretion rate. Thus, an interesting possibility to be explored in future work would include a model where the magnetic flux does not increase proportional to $\Phi \propto \sqrt{\dot{M}}$ but is truncated at a maximum value $\Phi \propto min(\sqrt \zeta \Phi_{0},\Phi_{max})$. This would cause the disk to transition from a MAD disk in the quiescent state to a SANE disk in the hard-intermediate state where, at least initially, the magnetic pressure is still dynamically important (e.g. \citealt{Begelman2017, Dexter2019, Lancova2019}). In upcoming work, we will employ ray-tracing calculations to compare both our existing models and future models featuring a truncated magnetic flux against multi-wavelength observations, which offer constraints on the truncation radius and the size/geometry of coronal structures in actual astrophysical systems (e.g. \citealt{Ingram2011, Plant2014, Fabian2014, Garcia2015, Kara2019}). 

There are several theoretical and observational arguments that support this `truncated flux' scenario. First, for systems to remain MAD during a 2-4 orders of magnitude increase in $\dot{M}$, they would need to advect $1-2$ orders of magnitude additional magnetic flux onto the Bh (e.g $\Phi \propto \sqrt{\dot{M}}$ in a MAD). Especially when the outer disk becomes geometrically thin it is unclear if this is physically possible since theoretical arguments suggest thin disks might not be able to advect magnetic flux loops (e.g. \citealt{Lubow1994}). Second, observations suggests that the disk truncation radius in the hard intermediate state appears (e.g. \citealt{Reis2010, Kara2019}) to be rather small ($r_t \lesssim 5 r_g$). This is inconsistent with recent GRMHD simulations which demonstrated that even when the disk only contained a factor $\sim 1.5$ of excess magnetic flux (above the MAD limit), this led to a truncation radius $r_t \sim 20 r_g$ (e.g. \citealt{Liska2018C,Liska2022}). Third, low-frequency quasi-periodic oscillations which are ubiquitous in the hard-intermediate state (e.g. \citealt{IngramMotta2020}), are most likely seeded by a precessing disk which tears of from a larger non-precessing disk (e.g. \citealt{Stella1998, Ingram2009, Ingram2016_QPO, Musoke2022}). This has been observed in radiative and non-radiative GRMHD simulations where a tilted thin accretion disk is threaded by a purely toroidal magnetic field (e.g. \citealt{Liska2020, Musoke2022, Liska2023}) and in similar GRMHD simulations where the accretion disk is threaded by a below saturation level vertical magnetic field (e.g. \citealt{Liska2019b}). However, there are no numerical simulations that have shown any disk tearing or precession where the disk is saturated by vertical magnetic flux (e.g. \citealt{Fragile2023}). The main problem is that for a disk to tear (and precess), the warping of space-time needs to substantially exceed the viscous torques holding the disk together (e.g. \citealt{Nixon2012, Nealon2015, Dogan2018, Dogan2020, Raj2021}). However, the viscous torques stemming from equipartition strength magnetic fields within the truncation radius might be too strong for a disk to tear.

While our simulations incorporate the effects of radiation and thermal decoupling between ions and electrons, they still rely on a rather simplistic heating prescription for electrons extracted from particle-in-cell models \citep{Rowan2017}. Since, absent any Coulomb collisions, the cooling rate in a given magnetic field will be determined by the temperature and density of the radiation emitting electrons, the radiative efficiency at lower accretion rates can become sensitive to the used heating prescription (e.g. \citealt{Chael2018}). For example, in our models roughly a fraction $\delta_{e} \sim 20-40 \%$ of the dissipation ends up in the electrons. If this electron heating fraction would be smaller/bigger, we expect that the radiative efficiency to drop/rise and the collapse to a two-phase medium to occur later/earlier. Similarly, other microphysical effects, typically not captured by the ideal MHD approximation, such as thermal conduction between the corona and disk (e.g. \citealt{Meyer1994, Liu1999, Meyer-Hofmeister2011, Cho2022, Bambic2024}), a non-unity magnetic Prandtl number (e.g. \citealt{Balbus2007}), could alter the transition rate to a two-phase medium.

In addition, it was recently demonstrated that the physics driving accretion in luminous black holes (e.g. with $L \gtrsim 0.01 L_{\rm Edd}$), which are misaligned with the black hole spin axis, is fundamentally different. Namely, dissipation of orbital energy is driven by nozzle shocks induced by strong warping \citep{Kaaz2022, Liska2023} instead of magneto-rotational instability (MRI) driven turbulence (e.g. \citealt{Balbus1991, Balbus1998}). These nozzle shocks form perpendicular to the line of nodes, where the disk's midplane intersects the equatorial plane of the black hole, and increase the radial speed of the gas by $2-3$ orders of magnitude in luminous systems that are substantially misaligned. This could, at a given accretion rate, lead to a decrease in the disk's density, potentially delaying the formation of a thin accretion disk. We expect to address outbursts of warped accretion disks using a simulation campaign performed at much higher resolutions.

Numerically, this article has also introduced a method to study outbursts by artificially rescaling the density as a function of time. This solves the issue that the physical processes in the outer disk that drive such drastic fluctuations in the mass accretion rate occur over timescales that are too long to simulate (real outbursts typically take weeks to months, while our simulations last for $t \sim 10-15s$). Future applications of this method might include (i) ultra-luminous accretion disks, which decay from super-Eddington to sub-Eddington accretion rates; (ii) the transition from the hard-intermediate state to the high-soft state, where the magnetic flux threading the black hole drops while the accretion rate remains constant; and (iii) the transition from the high-soft state to the quiescent state, characterised by a gradual drop in the mass accretion rate.

We also plan to address the structure and dynamics of the thin and truncated disks as we keep increasing the density scale with a dedicated simulation campaign performed at a much higher resolution. 

\section{Acknowledgements}
We thank Sera Markoff, Sasha Tchekhovskoy, and Ramesh Narayan for insightful discussions.  An award of computer time was provided by the Innovative and Novel Computational Impact on Theory and Experiment (INCITE) and ASCR Leadership Computing Challenge (ALCC) programs under awards PHY129 and AST178. This research used resources of the Oak Ridge Leadership Computing Facility, which is a DOE Office of Science User Facility supported under Contract DE-AC05-00OR22725. ML was supported by the John Harvard, ITC and NASA Hubble Fellowship Program fellowships. NK was supported by an NSF Graduate Research Fellowship. RE was supported by the NASA ATP grant numbers 21-ATP21-0077, NSF AST-1816420, and HST-GO-16173.001-A. KC was supported by the Black Hole Initiative (BHI) fellowship program. GM was supported by a Netherlands Research School for Astronomy (NOVA) Virtual Institute of Accretion (VIA) and the Canadian Institute for Theoretical Astrophysics (CITA) postdoctoral fellowships. 

\begin{figure}
    \centering
    \includegraphics[clip,trim=0.0cm 0.0cm 0.0cm 0.0cm,width=0.5\columnwidth]{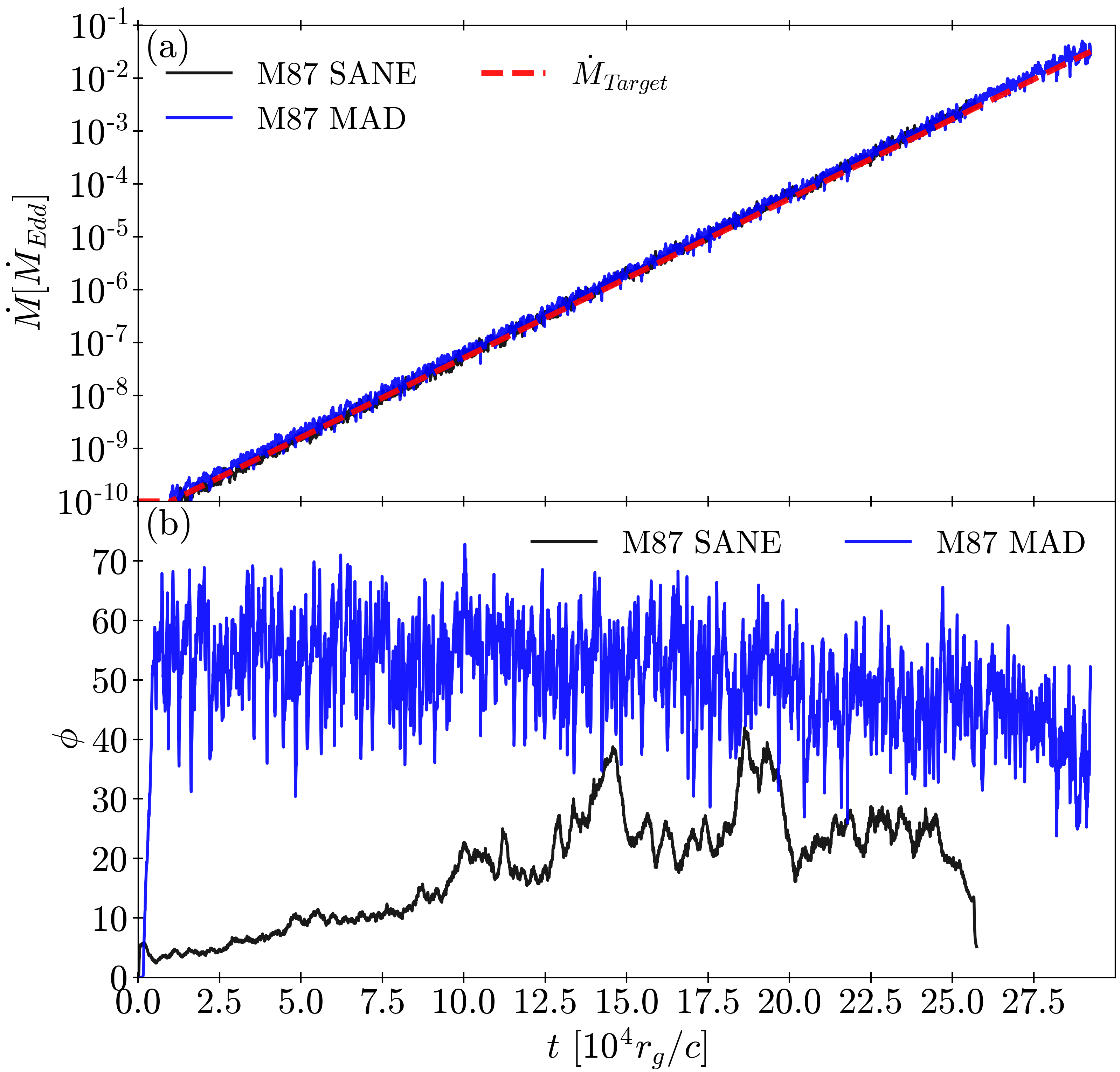}
    \vspace{-10pt}
    \caption{Same as figure \ref{fig:timeplot}, but for a $M=6.5 \times 10^9 M_{\odot}$ AGN.}
    \label{fig:timeplot_AGN}
\end{figure}

\appendix
   \vspace{-0.9cm}
 We present two radiative two-temperature GRMHD models (M87 SANE and M87 MAD) where we change the black hole mass from a typical BhXRB of $M_{\rm Bh}=10M_{\odot}$ to a large AGN such as M87 with $M_{\rm Bh}=6.5 \times 10^9 M_{\odot}$. Figures \ref{fig:timeplot_AGN}, \ref{fig:contourplot_AGN}, \ref{fig:streamplot_AGN},\ref{fig:mdotplot_AGN} and \ref{fig:coolplot_AGN} in the Appendix correspond to figures \ref{fig:timeplot}, \ref{fig:contourplot}, \ref{fig:streamplot_XRB}, \ref{fig:mdotplot} and \ref{fig:coolplot} in the main article. Interestingly, the evolution of our AGN models looks very similar to our BhXRB models. The most striking difference between BhXRBs and AGN is a slightly lower radiative efficiency at lower accretion rates, which can be explained by a weaker Synchrotron emission opacity coefficient (e.g. \citealt{Mckinney2017}) and is reflected in a longer emission time $t_{Em}$. In addition, after the plasma condenses into a two-phase medium, the temperature of the cold phase gas in AGN  ($T_e \sim 10^5K$) is much lower than in BhXRBs ($T_e \sim 10^7 K$). This is a well known fact in the analytic theory of radiatively efficient AGN accretion disks (e.g. \citealt{ss73, Novikov1973}), which are less dense and hence more radiation pressure dominated than their BhXRB analogues.

 \clearpage

\begin{figure*}
    \centering
    \includegraphics[width=1.0\linewidth,trim=0mm 0mm 0mm 0,clip]{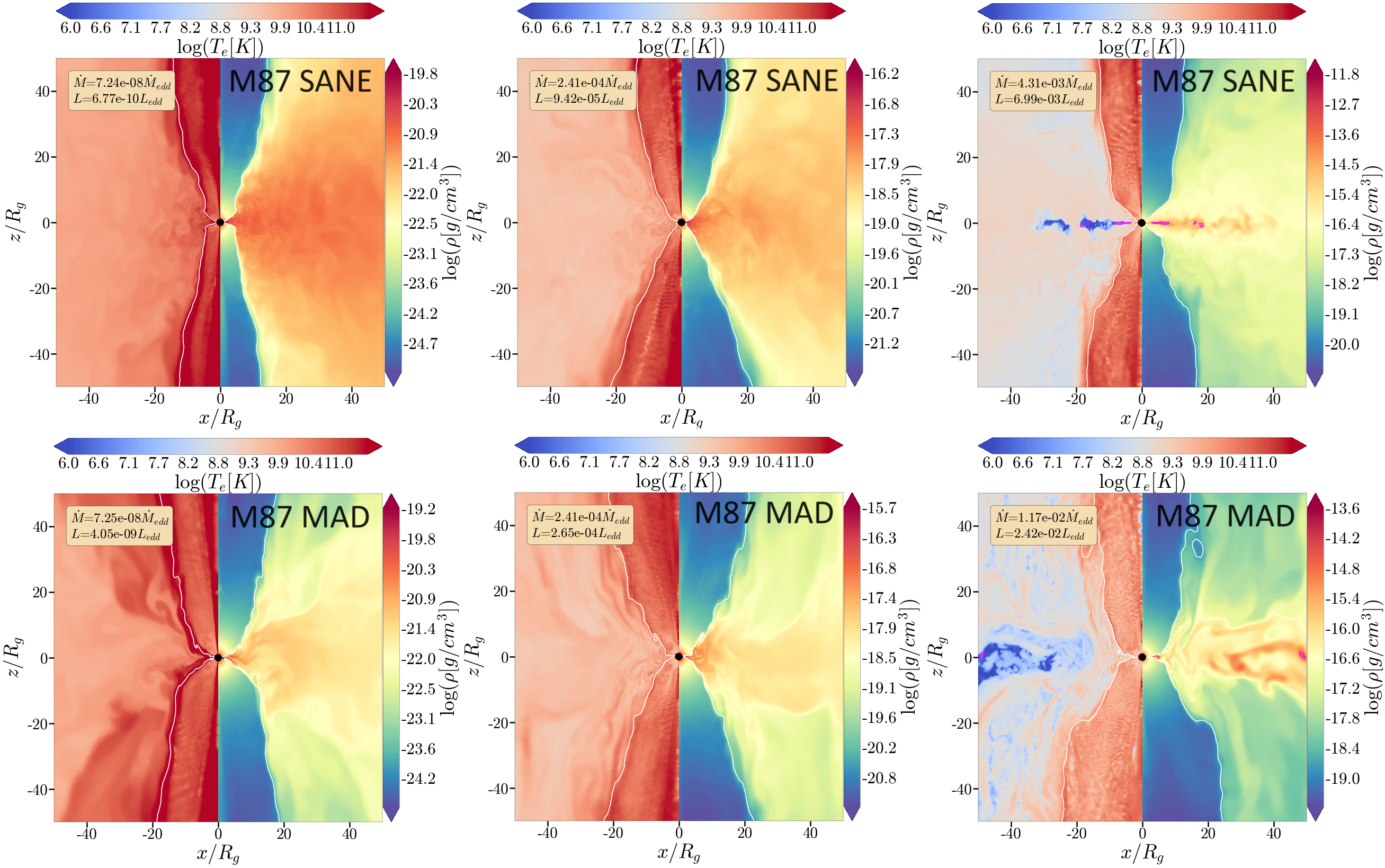}
    \caption{Same as figure \ref{fig:contourplot}, but for a $M=6.5 \times 10^9 M_{\odot}$ AGN.}
    \label{fig:contourplot_AGN}
\end{figure*}

\begin{figure}
    \centering
    \includegraphics[clip,trim=0.0cm 0.0cm 0.0cm 0.0cm,width=0.605\columnwidth]{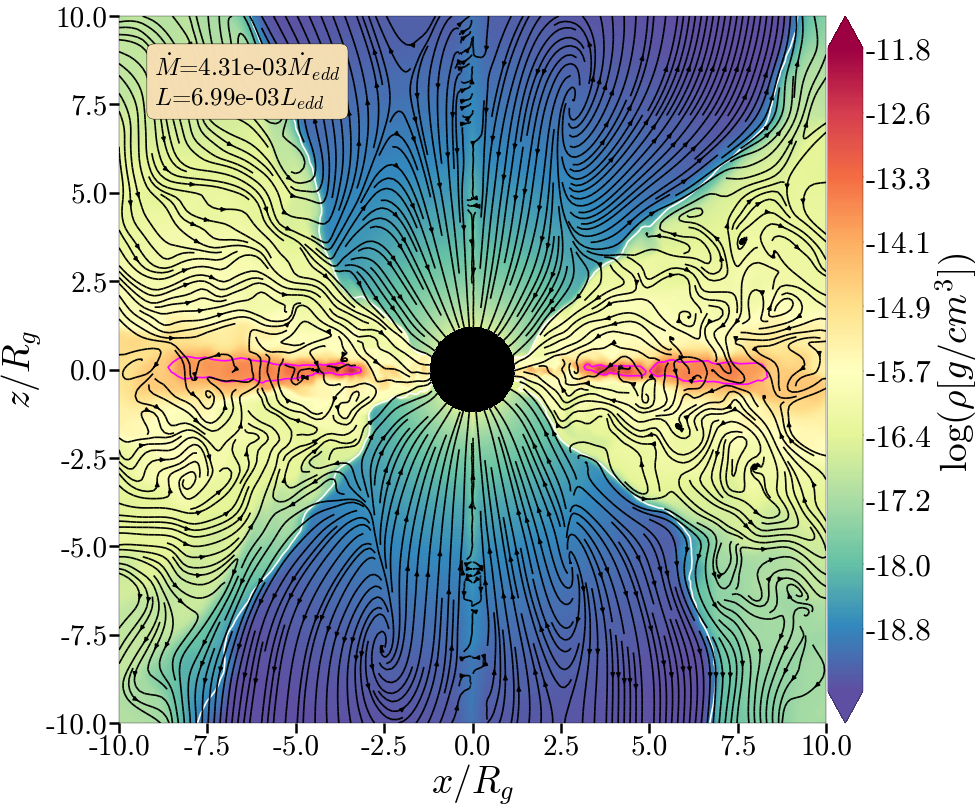}
    \vspace{-10pt}
    \caption{Same as figure \ref{fig:streamplot_XRB}, but for a $M=6.5 \times 10^9 M_{\odot}$ AGN.}
    \label{fig:streamplot_AGN}
\end{figure}

\begin{figure}
    \centering
    \includegraphics[clip,trim=0.0cm 0.0cm 0.0cm 0.0cm,width=0.605\columnwidth]{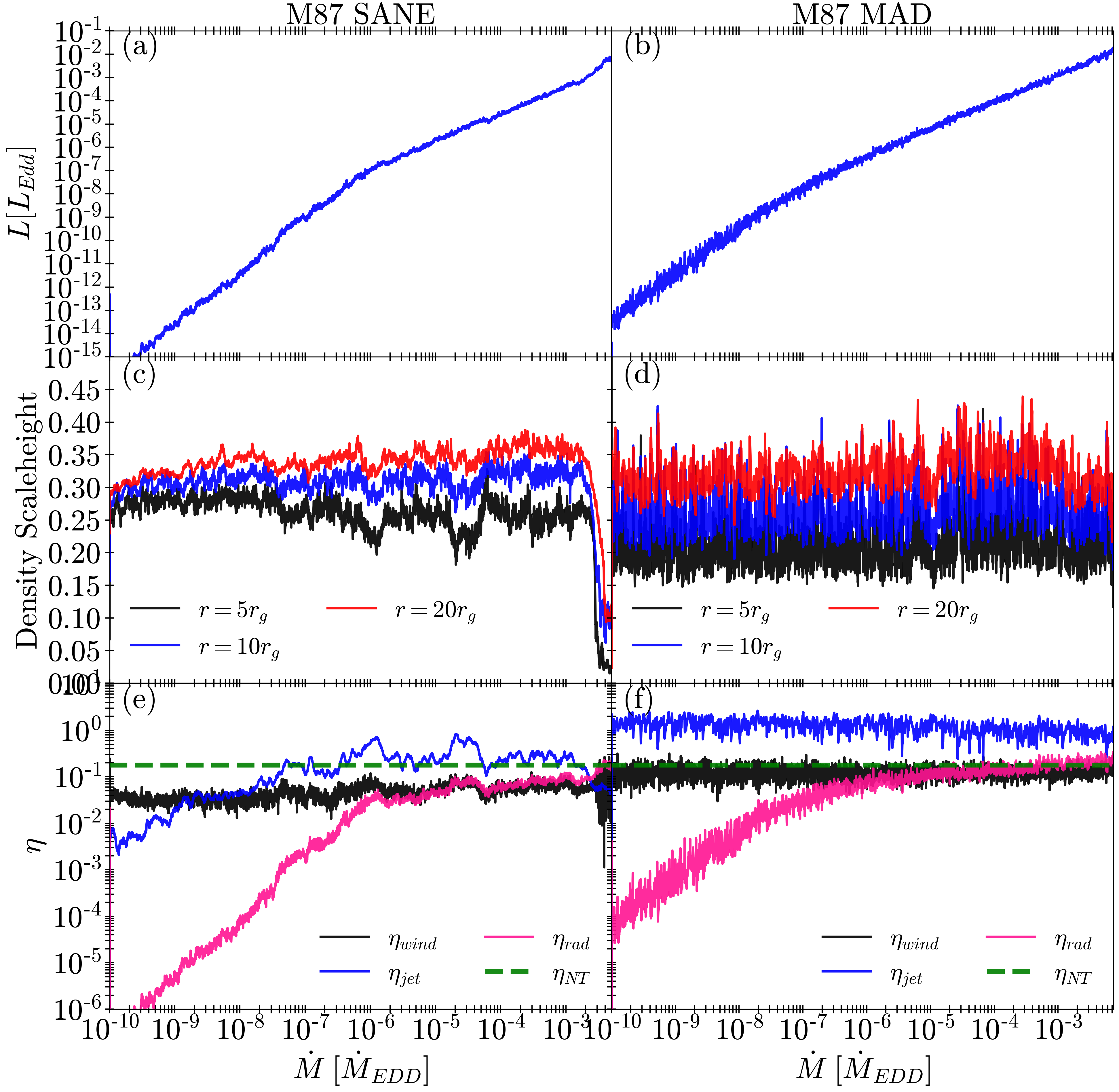}
    \vspace{-10pt}
    \caption{Same as figure \ref{fig:mdotplot}, but for a $M=6.5 \times 10^9 M_{\odot}$ AGN.}
    \label{fig:mdotplot_AGN}
\end{figure}

\begin{figure}
    \centering
    \includegraphics[clip,trim=0.0cm 0.0cm 0.0cm 0.0cm,width=0.605\columnwidth]{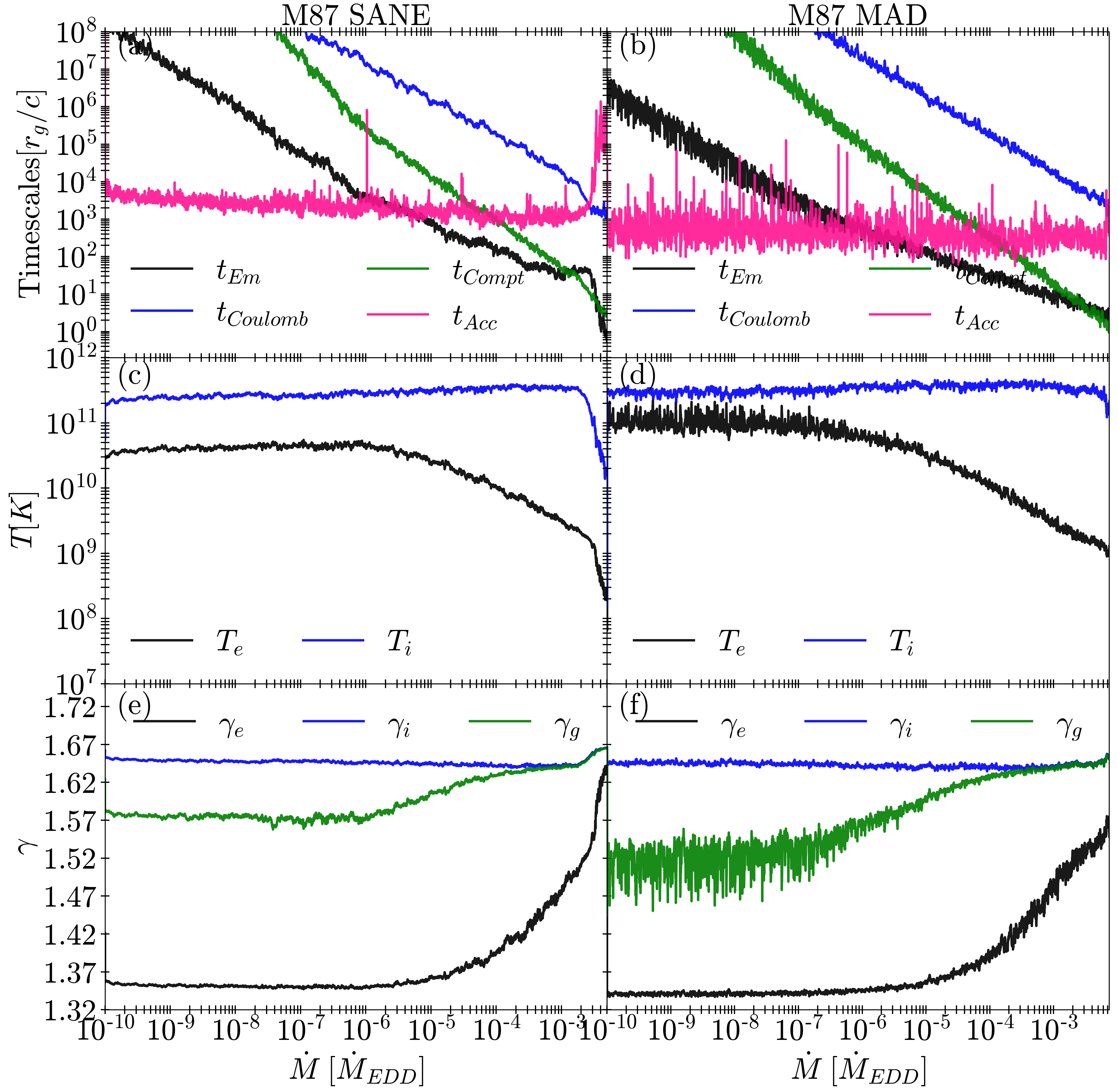}
    \vspace{-10pt}
    \caption{Same as figure \ref{fig:coolplot}, but for a $M=6.5 \times 10^9 M_{\odot}$ AGN.}
    \label{fig:coolplot_AGN}
\end{figure}

\newpage
\clearpage
\bibliography{new.bib,extra.bib}{}
\bibliographystyle{aasjournal}



\end{document}